\documentclass[printer]{aa}
\usepackage[english]{babel}
\usepackage{txfonts}
\usepackage{natbib}
\usepackage{aas_macros}
\usepackage{graphicx}
\def\xmm {\emph{XMM-Newton}}

\def\suz {\emph{Suzaku}}

\def\xte {\emph{RossiXTE}}
\def\int {\emph{INTEGRAL}}
\def\asca {\emph{ASCA}}
\def\hete {\emph{HETE-2}}
\def\src {SGR\,1806$-$20}

\def\flux {\mbox{erg cm$^{-2}$ s$^{-1}$}}
\def\lum {\mbox{erg s$^{-1}$}}
\def\nh {$N_{\rm H}$}
\begin{document}
\title{\src\ about two years after the giant flare: \suz, \xmm\ and \int\ observations}
\author{P. Esposito\inst{1,2}
\and S. Mereghetti\inst{2}
    \and A. Tiengo\inst{2}
    \and S. Zane\inst{3}
    \and R. Turolla\inst{4,3}
    \and D. G\"{o}tz\inst{5}
    \and \mbox{N. Rea}\inst{6,7}
    \and N. Kawai\inst{8}
    \and \mbox{M. Ueno}\inst{8}
    \and \mbox{G.L. Israel}\inst{9}
    \and \mbox{L. Stella}\inst{9}
    \and \mbox{M. Feroci}\inst{10}}
    
   \offprints{\mbox{Paolo Esposito, paoloesp@iasf-milano.inaf.it}}
   
\institute{Universit\`a degli Studi di Pavia, Dipartimento di Fisica Nucleare e Teorica and INFN-Pavia, via Bassi 6, I-27100 Pavia, Italy
  \and INAF - Istituto di Astrofisica Spaziale e Fisica Cosmica Milano, via Bassini 15, I-20133 Milano, Italy
\and Mullard Space Science Laboratory, University College London, Holmbury St. Mary, Dorking Surrey, RH5 6NT, UK
\and Universit\`a di Padova, Dipartimento di Fisica, via Marzolo 8, I-35131 Padova, Italy
\and CEA Saclay, DSM/DAPNIA/Service d'Astrophysique, F-91191 Gif-sur-Yvette, France\and SRON Netherlands Institute for Space Research, Sorbonnelaan 2, 3584 CA Utrecht, The Netherlands
\and University of Amsterdam, Astronomical Institute ``Anton Pannekoek'',
Kruislaan 403, 1098~SJ, Amsterdam, The Netherlands
\and Department of Physics, Tokyo Institute of Technology, 2-12-1 Ookayama, Meguro-ku, Tokyo 152-8551, Japan
\and INAF - Osservatorio Astronomico di Roma, via Frascati 33, I-00040 Monteporzio Catone, Italy
\and INAF - Istituto di Astrofisica Spaziale e Fisica Cosmica Roma, via del Fosso del Cavaliere 100, I-00133 Rome, Italy
}
\date{Received / Accepted}
\abstract{In December 2004, the soft gamma-ray repeater \src\ emitted the most powerful giant flare ever observed. This probably involved a large-scale rearrangement of the magnetosphere leading to observable variations in the properties of its X-ray emission.
Here we present the results of the first \suz\ observation of
\src, together with almost simultaneous observations with \xmm\
and \int. The source seems to have reached a state characterized
by a flux close to the pre-flare level and by a relatively soft
spectrum. Despite this, \src\ remained quite active also after the
giant flare, allowing us to study several short bursts observed by
\suz\ in the 1--100 keV range. We discuss the broad-band spectral
properties of \src, covering both persistent and bursting
emission, in the context  of the magnetar model, and consider its
recent theoretical developments.
 }
\titlerunning{X-ray observations of \src}
\authorrunning{P.~Esposito et al.}
\maketitle
\section{Introduction}
The four known Soft Gamma-ray Repeaters (SGRs) were discovered as
transient sources of high-energy photons; they emit sporadic and
short ($\sim$0.1 s) bursts of (relatively) soft gamma-rays with
luminosity $L\sim10^{40}$--$10^{41}$ \lum during periods of
activity, that are often broken by long intervals of quiescence.
Three ``giant'' flares with luminosity
\mbox{$\gtrsim$$10^{43}$ \lum} have also been observed to date, each
one from a different SGR: on March 5, 1979 from SGR\,0526--66 in
the Large Magellanic Cloud \citep{mazets79}, on August 27, 1998
from SGR\,1900+14 \citep{hurley99}, and on December 27, 2004 from
\src\ \citep{hurley05}. Persistent emission with $L\sim10^{35}$
\lum\ is also observed from SGRs in  the soft X--ray range
(\mbox{$<$10 keV}) and, for \src\ and SGR\,1900+14, also in the
hard X-ray range \citep{mgm05,gotz06}. In three cases, periodic
pulsations at a few seconds have been detected.
The bursts, the giant flares, the quiescent X-ray counterparts, and the pulsations have been interpreted in the framework of the magnetar model \citep[see][and references therein]{tlk02}. Magnetars are highly magnetized neutron stars with field strengths of \mbox{$10^{14}$--$10^{15}$ G}, larger than those of the majority of radio pulsars. The ultimate source of energy for the bursts and the quiescent emission is believed to be the ultra-strong magnetic field.\\
\indent \src\ was discovered in 1979 \citep{laros86,laros87} and its persistent X-ray counterpart was observed for the first time with the \asca\ satellite in 1993 \citep{murakami94}. A \xte\ observation led to the discovery of pulsations in the persistent emission with period $P\simeq7.47$ s and period derivative $\dot{P}\simeq2.6\times10^{-3}$ s  yr$^{-1}$ \citep{kouveliotou98}.
Under the assumption of pure magnetic dipole braking, these values imply a surface magnetic field strength of \mbox{$8\times10^{14}$ G}, strongly supporting the magnetar model.
Both the burst rate and the X-ray persistent emission of \src\ started increasing during 2003 and throughout 2004 \citep{mte05,tiengo05,met07,woods07}, culminating with the giant
flare of December 27, 2004, during which $\sim$$10^{47}$ erg were
released\footnote{Assuming isotropic luminosity and for a distance $d=15$ kpc \citep{corbel97,mcclure05}.
} \citep{hurley05,mereghetti05,terasawa05}.
This giant flare was exceptionally intense, $\sim$100 times more
energetic than those from  SGR\,0526--66 and SGR\,1900+14.
Observations with \xte\ unveiled, for the first time in an
isolated neutron star, rapid quasi-periodic oscillations in the
pulsating tail of the flare, likely related to global seismic oscillations on the neutron star surface \citep{ibs05}. The flare produced a
hard X-ray (\mbox{$>$80 keV}) afterglow lasting a few hours
\citep{mereghetti05,frederiks07} and a radio afterglow that faded in a few
days \citep{cameron05}. The small positional uncertainty of the
radio observations permitted to identify the likely IR counterpart
of the SGR \citep{kosugi05,israel05}. The fluxes observed in the
IR and gamma energy bands show a variability correlated
with that observed in the 2--10 keV energy range \citep{met07}.\\
\indent After the giant flare, the persistent X-ray flux of \src\ started to decrease from its
outburst level, and its X-ray spectrum to soften \citep{rea05,rea05_atel,met07,tiengo05,woods07}. A Similar flux decrease have been observed from its radio afterglow \citep{gaensler05,taylor05}
and its newly discovered IR counterpart \citep{israel05,rea05_atel,met07}.\\
\indent Here we present the results of the first \suz\ observation of \src, covering both persistent and bursting emission in the 1--100 keV energy band. We also report on the analysis of a simultaneous observation performed with \xmm\ and the  latest outcomes of the monitoring of \src\ with \int, comparing them with what is seen in the same energy ranges with \suz.
\section{\suz\ observation and analysis}
The \suz\ observation of \src\ started on September 09, 2006 at 23:13 UT and ended on September 11, at 04:01 UT. The \suz\ X-ray observatory \citep{mitsuda07} carries on board the XIS spectrometers \citep{koyama07} operating in the \mbox{0.2--12 keV} energy band, and the HXD collimated detector \citep{takahashi07}, which covers the \mbox{10--70 keV} energy range with PIN diodes and the \mbox{40--600 keV} with GSO scintillators.
Four X-ray telescopes with a spatial resolution (half-power diameter) of 2$'$ \citep[XRTs;][]{serlemitsos07} focus X-rays onto the four sensors (XIS 0, 1, 2, and 3) that constitute the XIS instrument. Each XIS contains 1024 by 1024 pixel rows covering a $18'\times18'$ field of view, and features an energy resolution of $\sim$140 eV at 6 keV. XIS 0, 2, and 3 are front-illuminated (FI) CCDs, while XIS 1 is a back-illuminated (BI) CCD, that features an enhanced soft X-ray response. The XRT/XIS combination yields effective area per detector of $\sim$330 cm$^2$ (FI) or $\sim$370 cm$^2$ (BI) at 1.5 keV, and of $\sim$160 cm$^2$ (FI) or $\sim$110 cm$^2$ (BI) at 8 keV.\\
\indent The 50 ks long observation was carried out with \src\ at
the ``HXD nominal'' pointing position. The XIS was operated in the
normal mode with the 3$\times$3 editing mode (time resolution of
\mbox{8 s}). The data sets were processed using the version 6.1 of
the \texttt{FTOOLS} package
and the most recent available calibration files available at the time the reduction was performed (November 2006). The XIS pipeline products were affected by an imperfect charge transfer inefficiency (CTI) correction, resulting in a systematically lower energy scale. The error has been corrected by applying the CTI correction tool \texttt{xispi} again with correct CTI parameters.\\
\indent For the XISs, source spectra were extracted from circular
regions with radii of  3$'$ centered at the position of \src,
while the background spectra from composite regions (far enough
from \src\ to prevent contamination by its photons). We screened
the XIS events based on standard criteria\footnote{See The Suzaku
Data Reduction Guide,\\
\texttt{http://suzaku.gsfc.nasa.gov/docs/suzaku/analysis/abc/}\,.}:
only events with GRADE 0, 2, 3, 4 and 6 were considered; the
\texttt{CLEANSIS} script was used to remove hot or flickering
pixels; data collected within 256 s of passage through the South
Atlantic Anomaly (SAA) were discarded; data were selected to be at
more than 5\degr\ in elevation above the Earth rim (20\degr\ above
the day Earth rim).
This resulted in a net exposure time of \mbox{46.4 ks} and about 57,000 net counts. The response matrices and effective area files were generated independently for each XIS with the tasks \texttt{xisrmfgen} and \texttt{xissimarfgen} (the ARF generator takes into account the level of hydrocarbon contamination on the optical blocking filter). The spectra were binned with \texttt{GRPPHA} following indications from the XIS Team; furthermore, the data were further rebinned to have at least 200 source events per bin.\\
\indent The HXD data were selected according to the following standard criteria: at least 500 s after the SAA passages, day and night Earth elevation angles each $\geqslant$5\degr, and geomagnetic cut-off rigidity to be at least 8 GeV c$^{-1}$. The exposure was corrected for the instrument dead time, for a net exposures of 48.4 ks in the PIN and 48.8 ks in the GSO. The HXD PIN and GSO instrumental background events were provided by the HXD Team (the instrumental background is due to events created by particles in the vicinity of the instrument). The source and background spectra (generated with the same good-time intervals) were both binned with \texttt{GRPPHA} following recommendations from the HXD Team.\\
\indent To study the properties of the persistent emission of \src, we cleaned the event list from bursts by applying intensity filters (with a negligible reduction of the net integration time). Spectral fits were performed using the \texttt{XSPEC} version 12.3 software. The analysis of the bursts is presented in Section \ref{burstingemission}.
\subsection{Results in the 1--10 keV energy range}\label{xisanalysis}
Owing to the high interstellar absorption, very few counts were detected from \src\ at low energies and thus we limited the spectral analysis to the \mbox{1.5--12 keV} energy range. \suz\ is placed in a near-circular orbit around the Earth with an orbital period of about 96 minutes. Due to the source occultation by the Earth in each orbit, the data-gathering required $\sim$1.2 days.
Apart from the bursts, no variability in the XIS light curves of \src\ was detected.
The 8 s time resolution of the XIS data does not allow to detect the $\sim$7.6 s pulsations.
We also investigated the possibility of spectral variability by splitting the observation in three intervals of equal duration, with negative results.\\
\indent We fit simultaneously the XIS spectra (with relative
normalization factors to account for the calibration uncertainties
between the four cameras, see Section \ref{xmm} for details)
adopting a power-law and a power-law plus blackbody model. The
reduced $\chi^2$  of the former fit, $\chi^2_r = 1.16$ for 283
degrees of freedom (d.o.f.), corresponding to a null hypothesis
probability of 0.03, is not completely satisfactory. The power-law
plus blackbody model provided a better fit, with $\chi^2_r = 0.98$
for 281 d.o.f. (null hypothesis probability = 0.6). The best fit
parameters are photon index \mbox{$\Gamma=1.8\pm0.1$}, blackbody
temperature $k_BT=0.49^{+0.08}_{-0.07}$ keV, and absorption \mbox{$N_{\rm
H}=7.1^{+0.6}_{-0.5}\times10^{22}$ $\rm cm^{-2}$} (Table \ref{fits}).
\begin{table*}
\begin{minipage}[h]{17cm}
\begin{center}
\caption{Summary of the spectral results for the persistent emission in the 1.5--12 $\rm{keV}$ energy range with \suz/XIS and \xmm/pn instruments. Errors are quoted at the 90\% confidence level for a single parameter.}
\label{fits}
\begin{tabular}{cccccccc}
\hline\hline
Model$^{\mathrm{a}}$ & Instruments &  \nh & $\Gamma$ & $k_B T$ & $R_{\rm{BB}}$$^{\mathrm{b}}$ & Flux$^{\mathrm{c}}$ & $\chi^{2}_{r}$ (d.o.f.)\\
 & & ($10^{22}$ $\rm cm^{-2}$) & & (keV) & (km) & ($10^{-11}$ \flux) & \\
 \hline
PL & XIS & $6.4\pm0.2$ & $2.03\pm0.04$ & -- & -- & $1.90\pm0.04$ & 1.16 (283) \\
& pn & $6.5\pm0.2$ & $1.78\pm0.05$ & -- & -- & $1.74\pm0.04$ & 1.25 (71) \\
 & XIS+pn & $6.5\pm0.1$ & $1.95\pm0.03$ & -- & -- & $1.73\pm0.03$ & 1.56 (356) \\
PL + BB & XIS &  $7.1^{+0.6}_{-0.5}$ & $1.8\pm0.1$ & $0.49^{+0.08}_{-0.07}$ & $5^{+9}_{-2}$ & $2.1\pm0.1$ & 0.98 (281) \\
& pn & $6.7^{+0.6}_{-0.7}$ & $1.6^{+0.1}_{-0.3}$ & $0.6^{+0.2}_{-0.1}$ & $2^{+3}_{-1}$ & $1.8^{+0.1}_{-0.2}$  & 1.09 (69) \\
 & XIS+pn &  $6.9\pm0.4$ & $1.6\pm0.1$ & $0.55\pm0.07$ & $3.7^{+1.6}_{-0.8}$ & $1.8\pm0.1$ & 1.33 (354) \\
\hline
\end{tabular}
\end{center}
\begin{list}{}{}
\item[$^{\mathrm{a}}$] Model applied in \texttt{XSPEC} notation: PL = \texttt{phabs*powerlaw} and PL+BB = \texttt{phabs*(powerlaw + bbodyrad)}. The abundances of \citet{anders89} are used
throughout.
\item[$^{\mathrm{b}}$] Radius at infinity assuming a distance of 15 kpc.
\item[$^{\mathrm{c}}$] Flux in the 2--10 keV range, corrected for the absorption.
\end{list}
\end{minipage}
\end{table*}
The presence of the blackbody component is consistent
with the findings of deeper \xmm\ observations \citep[][ see also
Section \ref{xmm}]{met07}.
\subsection{Results in the 10--100 keV energy range and broad-band spectral results}\label{hxd-analysis}
\begin{figure}[h!]
\resizebox{\hsize}{!}{\includegraphics[angle=-90]{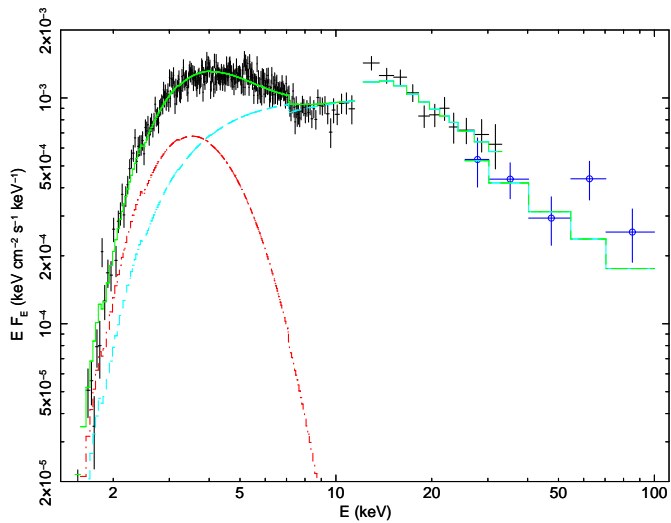}}
\caption{Broad-band \suz\ spectrum of \src\ (see the online edition of the article for a color version of this figure). The \suz's XIS\,023 and HXD-PIN data (in black) are fit with the broken power-law (light blue dashed line) plus blackbody  (red dot-dashed line) model. The XIS\,1 data are not shown for clarity. We also plotted the \int\ data (see Section\,\ref{igr-analysis}) using the blu circle marks.}
\label{suz-broad}
\end{figure}

The advantages of \suz/HXD over previous non imaging instruments are its small field of view ($34'$$\times$$34'$ FWHM below $\sim$100 keV) and a low instrumental background. The images obtained from \int\ very deep exposures do not show contaminating point sources within the HXD field of view \citep[see][Fig.\,1]{mgm05} and no bright and hard X-ray sources below 10 keV have been found either in the \asca\ Galactic Plane survey \citep{sugizaki01} or in the SIMBAD database\footnote{See \texttt{http://simbad.u-strasbg.fr/simbad/}\,.}. However, given that \src\ lies at low Galactic latitude and longitude ($b\simeq0\degr$ and $l\simeq10\degr$), the study of its emission in the hard X\,/\,soft gamma-ray band is complicated by the presence of the diffuse emission from the Galactic Ridge \citep[see][and references therein]{lebrun04}.\\
\indent After standard data processing, a positive flux possibly
associated with \src\ is detected in the HXD-PIN data up to
$\sim$40 keV (apart from the bursts, no significant emission is
detected in the GSO data). The instrumental background counts
obtained by simulations based on the present knowledge of HXD
in-orbit performances, are about 70\% of the $\sim$26,400 total
counts in the 12--40 keV band. To estimate the cosmic X-ray
background level in the HXD-PIN band we took the spectrum reported
in \citet{gruber99}, of the form
$7.877\,E^{-0.29}e^{-E/41.13\,\rm{keV}}$ keV keV$^{-1}$ cm$^{-2}$
s$^{-1}$ sr$^{-1}$. 
To model the Galactic Ridge emission we used the spectrum
reported in \citet{valinia98} for their R1 region (Central Ridge:
\mbox{$-1.5\degr<b<1.5\degr$} and $-45\degr<l<45\degr$), where the
SGR is located: a power-law with photon index $\Gamma=2.1$ and
surface brightness of \mbox{$4.9\times10^{-8}$ erg  cm$^{-2}$
s$^{-1}$ sr$^{-1}$} in the 10--35 keV band.\\
\indent However, the R1 region is wide and, since the Galactic Ridge emission strongly varies with latitude and longitude \citep{lebrun04}, this estimate of the Galactic Ridge contribution to the background could be severely inaccurate.
Therefore, we analyzed a 43 ks long pointing carried out with
\suz\ on April 07, 2007 at the coordinates $b\simeq0\degr$ and
$l\simeq8\degr$, to provide a background field for the observation
of the TeV source HESS\,J1804$-$216 \citep[reported
in][]{bamba07}. After a data processing  and instrumental
background subtraction performed as described above, we fit the
spectrum with a power-law  with photon index fixed to 2.1. The
ratio between the measured power-law normalization and the one
reported for the average spectrum in \citet{valinia98} is
$0.9\pm0.3$ (90\% confidence level).
So, these data indicate that the emission from the Galactic Ridge at a
longitude closer to that of \src\ is not significantly different from the average
value in the R1 region.
We remark that this result relies on the accuracy of the instrumental background estimate, at present $\lesssim$5\% \citep{kokubun07}. In fact, using a  5\% higher background we obtain a nearly null Galactic Ridge emission, whereas a 5\% smaller background yields a Galactic Ridge flux $2.0\pm0.4$ times higher than that of \citet{valinia98}.\\
\indent We note that to account for the whole signal detected in the HXD-PIN instrument, the Galactic Ridge emission in the HXD field of view should be $\sim$7 times higher than that reported by \citet{valinia98}. This seems very unlikely to us and therefore we consider significant the detection of \src. However, both its spectral shape and flux are subject to the uncertainty reflecting the coarse knowledge of the Galactic Ridge contribution to the background.\\
\indent Including the cosmic diffuse and Galactic Ridge emission as fixed components, we fitted the HXD-PIN spectrum in the \mbox{12--40} keV band to a  power-law model. The best-fit  parameters are $\Gamma=2.0\pm0.2$ and flux in the \mbox{20--60 keV} of \mbox{$(3.0\pm0.5)\times10^{-11}$ \flux} ($\chi^2_r=0.90$ for 10 d.o.f.). The source, with a net count rate of \mbox{$\sim$0.11 counts s$^{-1}$}, accounts for $\sim$70\% of the counts remained after the subtraction of the instrumental background. Varying the assumed  instrumental background by $\pm$5\% we obtain best-fit fluxes of \mbox{$(3.7\pm0.5)\times10^{-11}$ \flux} and \mbox{$(2.4\pm0.4)\times10^{-11}$ \flux}.\\
\indent Fitting together the HXD-PIN and XIS\footnote{For the broad-band analysis we added the spectra of the three FI CCDs (XIS\,0, 2, and 3) using the \texttt{FTOOL matpha}. We generated the instrumental responses by summing the redistribution matrices and the effective area files using the \texttt{\texttt{FTOOLS}} \texttt{addrmf} and \texttt{addarf}.} spectra, we find that the HXD-PIN data must be scaled downward by a factor of $\sim$2 to be consistent with the parameters derived in the 1.5--12 keV energy range. This scaling factor is unacceptably large, since the uncertainty in the relative calibration of the two instruments in the energy band considered here is of $\lesssim$20\% \citep{kokubun07}. To better reproduce the broad-band spectrum we tried a broken power-law plus blackbody model, with a normalization factor between the instruments kept at $<$1.2. We find an acceptable fit ($\chi^2_r = 1.09$ for 354 d.o.f.; null hypothesis probability = 0.13) with the photon index changing from $1.0\pm0.1$ below the break at $16\pm2$ keV to $2.2^{+0.4}_{-0.2}$ above it, $k_B T=0.8\pm0.1$ keV, $R_{\rm{BB}}= 2.5^{+0.4}_{-0.3}$ km (at \mbox{15 kpc}), and $N_{\rm H}=5.6^{+0.3}_{-0.4}\times10^{22}$ $\rm cm^{-2}$ (see Fig.\,\ref{suz-broad}). The corresponding 2--10 keV and 20--60 keV unabsorbed fluxes are \mbox{$\sim$$2 \times 10^{-11}$ \flux} and $\sim$$3 \times 10^{-11}$ \flux, respectively. The normalization factor assumed the value of 1.195, very close to the allowed maximum.
\\
\indent We did not find a significant pulsation in the HXD-PIN data (time resolution of \mbox{61 $\mu$s}). However, given the low signal-to-noise ratio, we do not expect to detect a clear signal if the pulsed fraction is $\sim$10\% (with a sinusoidal profile) as in the \mbox{2--10 keV} energy range, or smaller. By folding the HXD-PIN light curve on the \src\ pulsation period measured in simultaneous \xmm\ data (Section \ref{xmm}) and fitting it with a sinusoid, we determine a 3\,$\sigma$ confidence level upper limit of $\approx$20\% on the amplitude of a sinusoidal modulation (Galactic Ridge emission subtracted). This upper limit is consistent with the preliminary results obtained with \int\ in the \mbox{20--60 keV} energy range (G\"{o}tz et al. 2007, in preparation).

\subsection{Analysis of the bursts}\label{burstingemission}
\begin{figure*}[t!]
\centering
\includegraphics[width=18cm,angle=0]{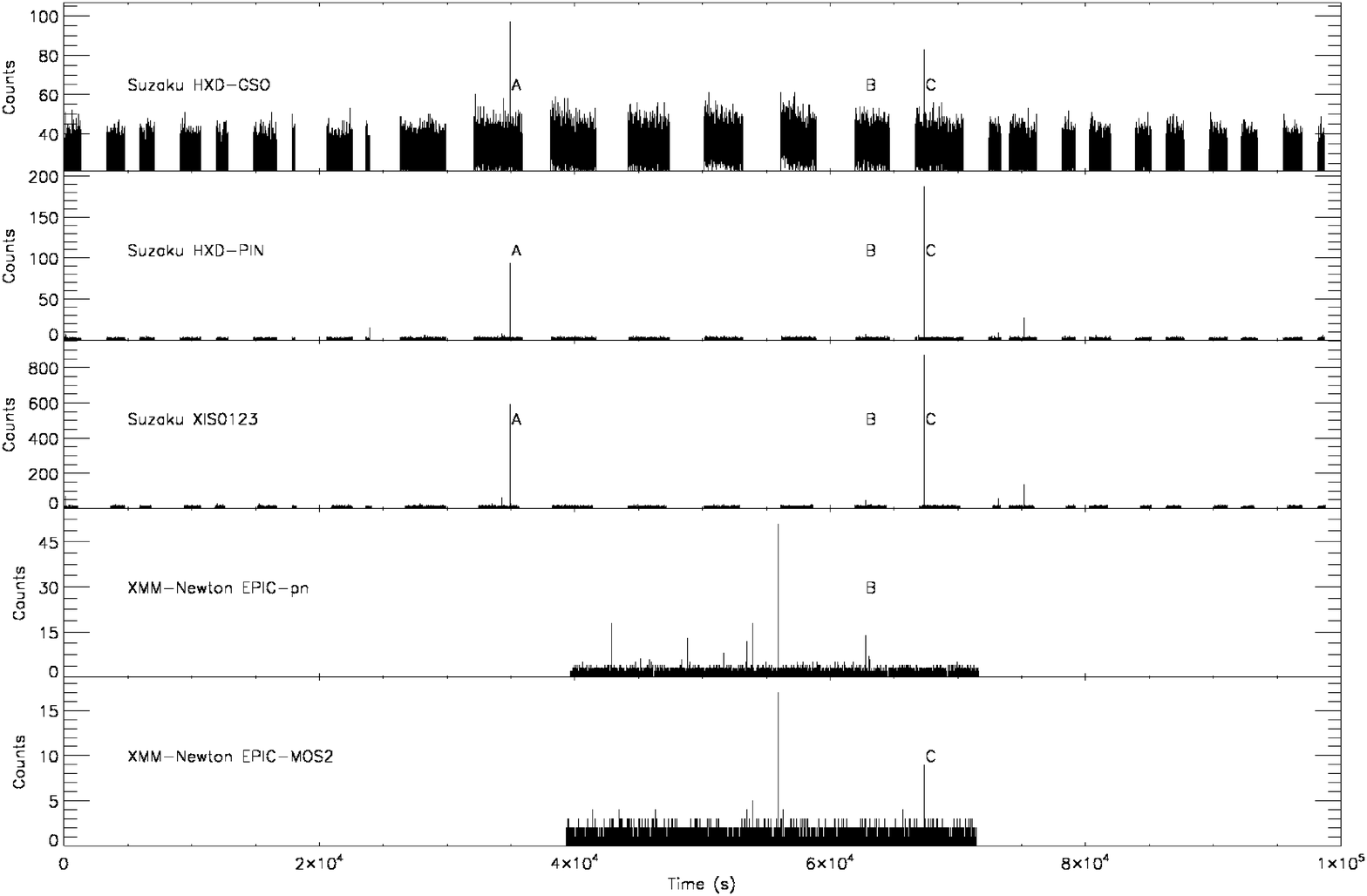}
\caption{Temporal coverage of the \suz\ and \xmm\ data (source plus background counts). Several bursts have been detected in the various instruments, but the brigh burst C is not visible in \xmm/pn because it is too strong and saturated the detector. Despite this, the burst rate in the \xmm/pn is higher than in in the \suz/XIS, with $0.5\pm0.2$ burst ks$^{-1}$ versus \mbox{$0.15\pm0.06$ burst ks$^{-1}$}. The difference is likely due to faint bursts undetected in the XIS owing to the long frame-time (8 s). The \suz\ arrival times of the photons were barycentered to the Solar System  using the task \texttt{aebarycen}. The comparison between the \suz/HXD and \xmm/EPIC times of bursts B and C proves a relative timing accuracy better than \mbox{0.2 s}.}
\label{suzxmm}
\end{figure*}

\begin{figure}[h!]
\resizebox{\hsize}{!}{\includegraphics[angle=0]{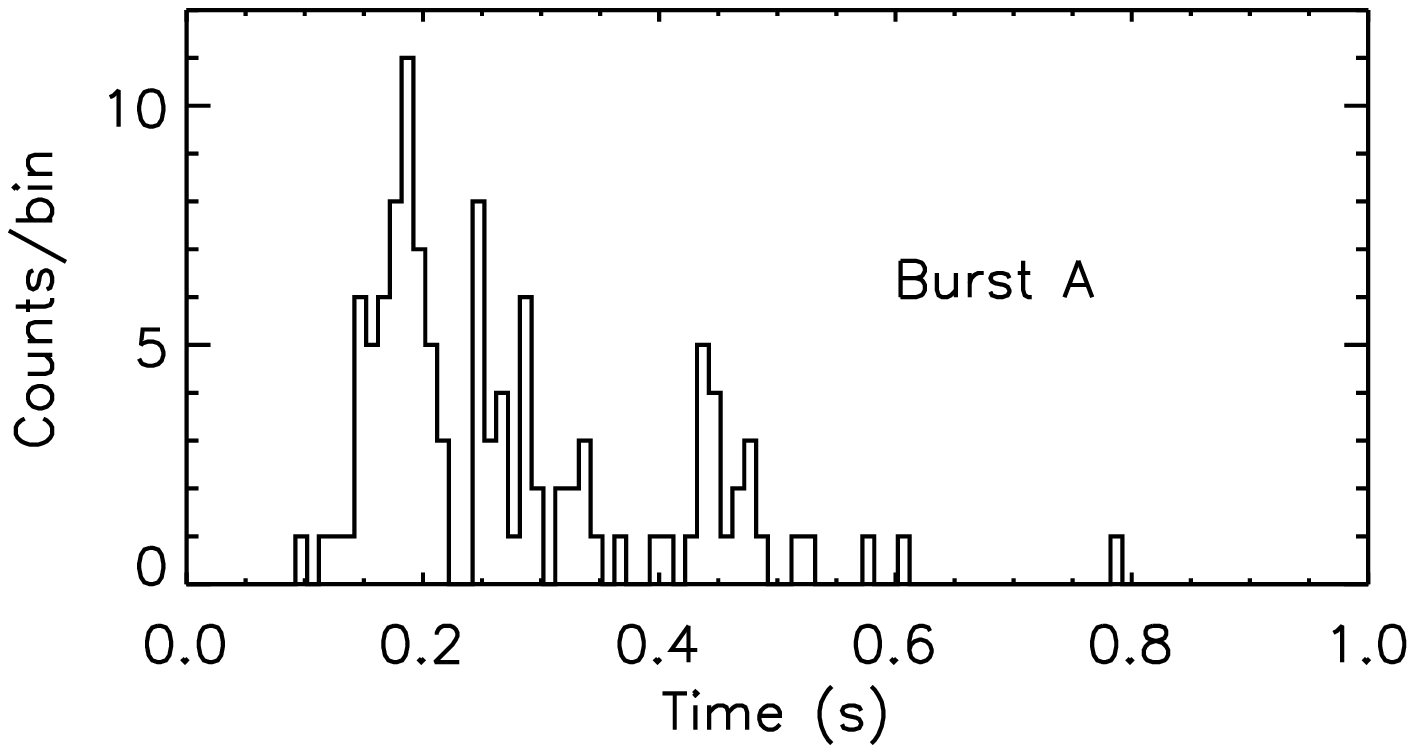}}

\vspace{-.5cm}

\resizebox{\hsize}{!}{\includegraphics[angle=0]{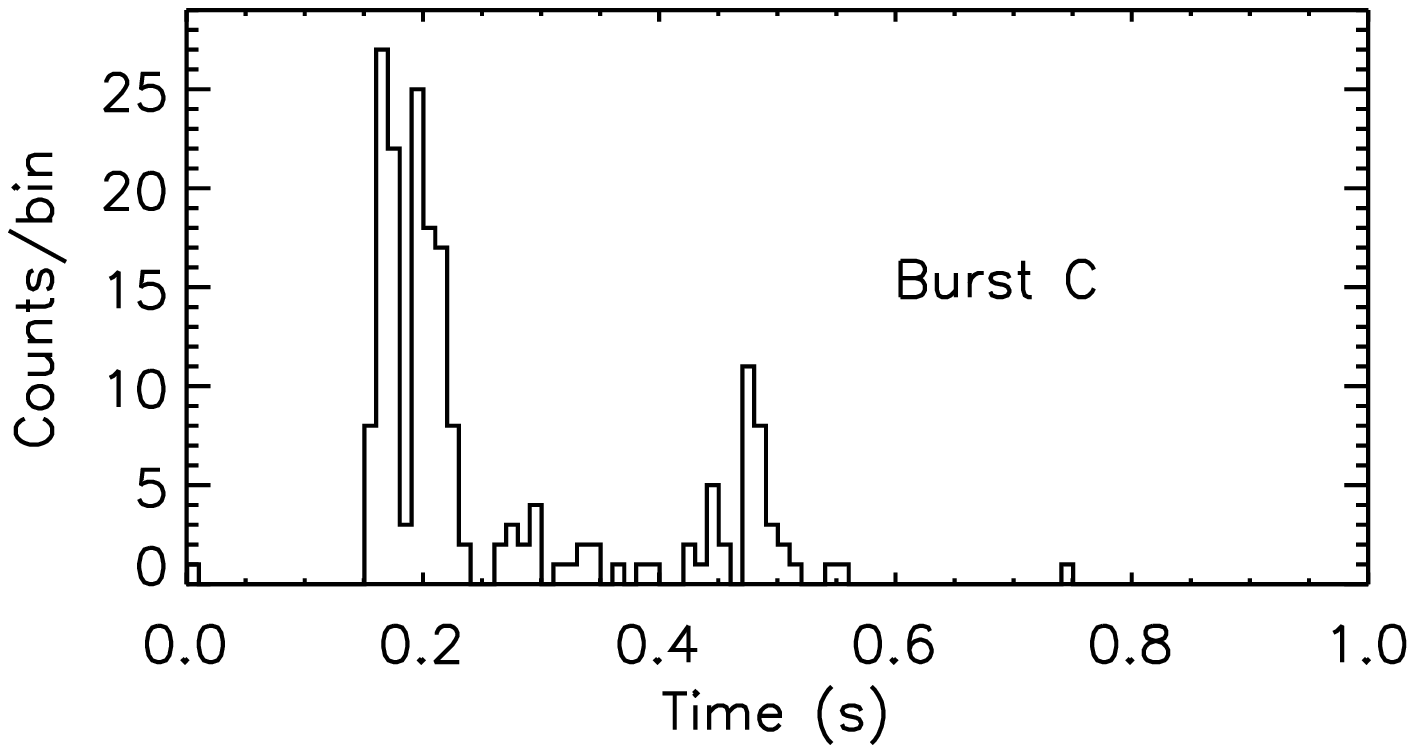}}
\caption{HXD-PIN light curves of the bursts A and C in the  energy range 12--50 keV. The origins of the time axis are arbitrary and the time bin size is 10 ms.}
\label{burstsab}
\end{figure}

\begin{figure*}[t]
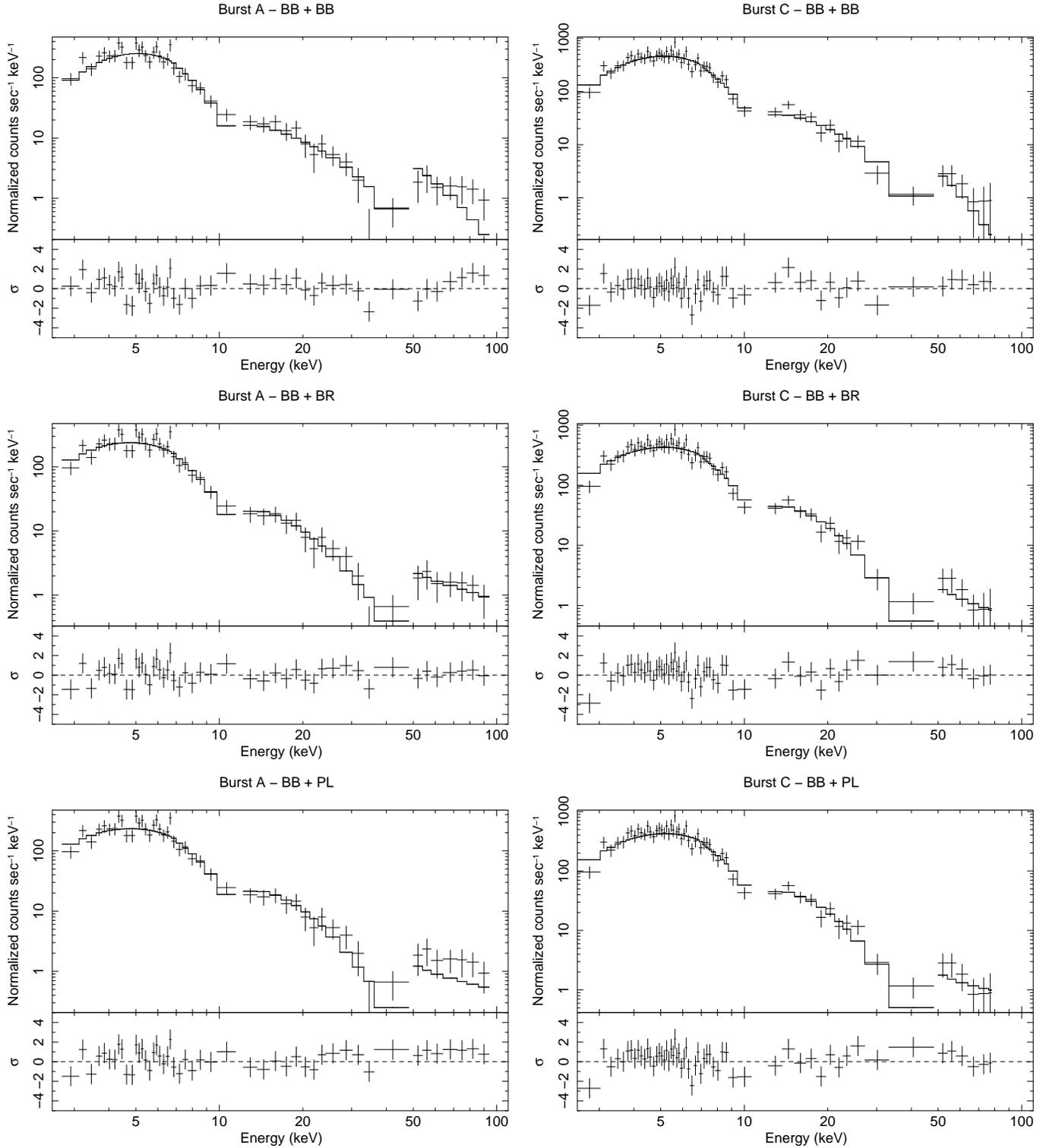

\resizebox{\hsize}{!}{\includegraphics[angle=-90]{./8562fig4a.eps}\hspace{1cm}\includegraphics[angle=-90]{./8562fig4b.eps}}

\vspace{0.3cm}

\resizebox{\hsize}{!}{\includegraphics[angle=-90]{./8562fig4c.eps}\hspace{1cm}\includegraphics[angle=-90]{./8562fig4d.eps}}

\vspace{0.3cm}

\resizebox{\hsize}{!}{\includegraphics[angle=-90]{./8562fig4e.eps}\hspace{1cm}\includegraphics[angle=-90]{./8562fig4f.eps}}
\caption{\suz\ broad-band spectra and residuals of the bursts A and C (XIS 1.5--12 keV, HXD-PIN 12--50 keV and HXD-GSO 50--100 keV). The model adopted is indicated on each panel (see also Table \ref{bfits})}
\label{burstfits}
\end{figure*}

The \suz\ (XIS, PIN, and GSO) light curves show many short bursts (Fig.\,\ref{suzxmm}).
To obtain significant constraints on spectral fit parameters, we considered only the events with more than 50 counts in the HXD-PIN band. Thus, we selected only two bursts, indicated by the labels A and C in Fig.\,\ref{suzxmm} and shown in Fig.\,\ref{burstsab}.
\\
\indent Since the frame-time of the XIS instruments (8 s) is much larger than the burst duration, we measured the duration of the bursts in the high time resolution HXD-PIN light curve (\mbox{$\sim$0.5 s} for burst A and $\sim$0.4 s for burst C) and we set the integration time of the XIS spectra to these values\footnote{During the two 8 s frames containing bursts A and C, less than 1.5\% of the total XIS counts can be attributed to the persistent source plus background emission. Thus, the effect of this emission on the XIS spectra of the bursts is negligible.}. The threshold above which the photon pileup becomes significant in the XIS cameras is $\sim$100 counts frame$^{-1}$ for a point source, and both bursts had a count rate above this value. After a careful analysis of the distribution of the counts in the pixels, we excluded a circular region with radius of 15$''$ at the image center from the event extraction region to minimize the pileup effects. To increase the statistics in the spectra we added the four XIS spectra using the \texttt{FTOOL matpha}. We calculated the corresponding instrumental responses by summing the redistribution matrices and the new effective area files for the annular regions using the \texttt{\texttt{FTOOLS}} \texttt{addrmf} and \texttt{addarf}. XIS and HXD spectra were rebinned to have a minimum number of 20 counts in each bin. For the background subtraction, under the usual working hypothesis that the bursting emission is present on top of the quiescent one, we used the spectra of the whole observation, cleaned from all the bursts.\\
\indent The column density of neutral absorbing gas along the line of sight is a critical parameter for the spectral fitting. The most precise measurement of this parameter comes from modeling the persistent X-ray emission: for \src\ the column density is consistently measured at \mbox{$N_{\rm H}\simeq6.5\times10^{22}$ $\rm cm^{-2}$} \citep[see Table\,\ref{fits} and][]{met07}. To test different spectral models we therefore decided to fix the equivalent hydrogen column at the value measured for the persistent source before and after the bursts.
Spectral fits with power-law or thermal bremsstrahlung models, with the \nh\ value fixed at \mbox{$6.5\times10^{22}$ $\rm cm^{-2}$}, yield unacceptable $\chi^2$ values. Good fits with either the power-law or bremsstrahlung, if the \nh\ is left free to vary, require large absorption: \nh\ values of \mbox{$1.7\times10^{23}$ $\rm cm^{-2}$} and \mbox{$1.4\times10^{23}$ $\rm cm^{-2}$}, respectively. Similar results were obtained by \citet{fenimore94} and by \citet{feroci04} for the soft gamma-ray repeater SGR\,1900+14. A single blackbody spectrum gives a formally acceptable fit in the soft range, but it severely underestimates the observed flux at higher photon energies. Another possible thermal model is the sum of two blackbodies, as used by \citet{olive04} and \citet{feroci04} to fit bursts from SGR\,1900+14. We found that this two components model provided good fits to the bursts, with the parameters summarized in Table \ref{bfits}. We tried other types of spectral models, obtaining almost equally good fits with either a blackbody plus power-law model or a blackbody plus thermal bremsstrahlung model (see Table \ref{bfits} and Fig.\,\ref{burstfits}).\\
\begin{table*}
\begin{minipage}[h]{17cm}
\begin{center}
\caption{Burst spectral parameters for the two-blackbody model (assuming for the absorption the value of $6.5\times10^{22}$ $\rm cm^{-2}$). Errors are quoted at the 90\% confidence level for a single parameter.}
\label{bfits}
\begin{tabular}{cccccccccc}
\hline\hline
Burst & Model$^{\mathrm{a}}$ & $k_B T_1$ & $R_{\rm{BB}\,1}$$^{\mathrm{b}}$ & $k_B T_2$ & $R_{\rm{BB}\,2}$$^{\mathrm{b}}$ & $k_B T_{\rm{BR}}$ & $\Gamma$ & Flux$^{\mathrm{c}}$ & $\chi^{2}_{r}$ (d.o.f.)\\
 & & (keV) & (km) & (keV) & (km) & (MeV) & & (\flux) & \\
\hline
A & BB + BB & $2.3^{+0.6}_{-0.4}$ & $15^{+5}_{-4}$ & $10\pm2$ & $1.5^{+0.7}_{-0.6}$ & -- & -- & $1.3\times10^{-7}$ & 1.15 (47) \\
  & BB + BR & $4.6^{+1.2}_{-0.8}$ & $5\pm2$ & -- & -- & $0.10^{+0.06}_{-0.03}$ & -- &$1.2\times10^{-7}$ & 1.08 (47) \\
  & BB + PL & $4.5^{+0.8}_{-0.7}$ & $5^{+2}_{-1}$ & -- & -- & -- & $1.5^{+0.2}_{-0.1}$ & $1.1\times10^{-7}$ & 1.19 (47) \\
C & BB + BB  & $2.2^{+0.5}_{-0.4}$ & $20^{+5}_{-4}$ & $7\pm1$ & $4\pm1$ & -- & -- & $2.1\times10^{-7}$ & 0.95 (63) \\
 & BB + BR  & $4.2^{+0.5}_{-0.4}$ & $9\pm2$ & -- & -- & $0.11^{+0.29}_{-0.06}$ & -- & $2.0\times10^{-7}$ & 1.09 (63) \\
 & BB + PL & $4.2\pm0.4$ & $9\pm2$ & -- & -- & -- & $1.4\pm0.2 $ & $2.0\times10^{-7}$ & 1.12 (63) \\
\hline
\end{tabular}
\end{center}
\begin{list}{}{}
\item[$^{\mathrm{a}}$] Model applied in \texttt{XSPEC} notation: BB + BB = \texttt{phabs*(bbodyrad + bbodyrad)}, BB + BR = \texttt{phabs*(bbodyrad + bremss)}, and \mbox{BB + PL = \texttt{phabs*(bbodyrad + powerlaw)}}.
\item[$^{\mathrm{b}}$] Radius at infinity assuming a distance of 15 kpc.
\item[$^{\mathrm{c}}$] Flux in the 2--100 keV range, corrected for the absorption.
\end{list}
\end{minipage}
\end{table*}

\section{\xmm\ observation and comparison with \suz}\label{xmm}
The \xmm\ observation of \src\ started on September 10, 2006 at 10:11 UT and ended at 19:04 UT. It is therefore simultaneous to part of the \suz\ observation, as shown in Fig.\,\ref{suzxmm}.
Here we present the analysis of the data collected with the EPIC instrument, which consists of two MOS \citep{turner01} and one pn \citep{struder01} cameras sensitive to photons with energies between 0.1 and 15 keV. The pn was operated in Small Window mode (time resolution 6 ms), while the MOS\,1 unit was in Timing mode (time resolution 1.5 ms) and the MOS2 in Full Frame mode (time resolution 2.6 s); all the detectors mounted the medium thickness filter. The data reduction was performed following the procedure described in \citet{tiengo05} but using the XMM-Newton Science Analysis Software (\texttt{SAS}) version 7.0.\\
\indent By an inspection of the \xmm\ lightcurves we found several bursts (see Fig.\,\ref{suzxmm}), but the \xmm\ data do not provide significant improvement on the results obtained from the \suz\ data (Section \ref{burstingemission}). In fact burst A occurred before the start of the \xmm\ exposure, while burst C was too bright for the pn (the two MOSs registered few counts only), saturating the instrument telemetry\footnote{If the count rate in the pn is higher than the telemetry limit (\mbox{$\sim$600 counts s$^{-1}$} for the imaging modes), then the so-called ``counting mode'' is triggered and for some time the science data are lost.}. Only burst B was observed by both satellites (see Fig.\,\ref{suzxmm}), but it was too faint for a spectral study. To obtain the results presented in this section, we excluded the bursts from the analysis by applying intensity filters.
A further cleaning was necessary because of the presence of soft proton flares during the observation. On the whole, the net exposure time was reduced from $\sim$22.3 ks to $\sim$21.6 ks for the pn detector.\\
\indent We fit in the 1.5--12 keV range the spectrum obtained with the pn camera adopting the same models used for the \suz\ spectral analysis. We again found that the power-law plus blackbody model provides a slightly better fit (see Table \ref{fits}); the best-fit parameters are \mbox{$\Gamma=1.6$}, \mbox{$k_BT=0.6$ keV}, and \mbox{$N_{\rm H}=6.7\times10^{22}$ $\rm cm^{-2}$}, with a $\chi^2_r$ of 1.09 for 69 d.o.f.. The corresponding luminosity in the \mbox{2--10 keV} band is \mbox{$5\times10^{35}d^2_{15}$ \lum}, where we indicate with $d_{\rm{N}}$ the distance in units of N kpc. This luminosity is slightly higher than that of the previous \xmm\ observation performed on April 4, 2006 \citep{met07}, as also supported by a simple comparison of the pn net count rates (\mbox{$0.993\pm0.007$ counts s$^{-1}$} with respect to \mbox{$0.946\pm0.007$ counts s$^{-1}$}), while the other spectral parameters are consistent. We also analyzed in a similar way the spectra obtained with the MOS cameras, finding results consistent with the pn ones.\\
\indent Given the simultaneity of the \xmm\ and \suz\ observations, we have tried to fit together the \xmm/pn and \suz/XIS spectra, with a normalization factor to account for the uncertainty in the absolute flux estimate of the different instruments.
In Table \ref{fits} we report the results of such analysis using either a power-law or a power-law plus blackbody model. We note that the $\chi^2$ values are unacceptable. Since no intrinsic spectral variability during the non-coincident exposure windows of the two satellites is expected (the bursts have been removed in both datasets), this simultaneous fit can be used to evaluate the cross-calibration discrepancies between the \xmm/pn and the four \suz/XIS detectors. Fixing the \xmm/pn normalization factor to 1 and linking all the other parameters, we derive the normalization factors reported in Table \ref{crossc}, either for the whole energy range or restricting the fit to three energy bands. These values show that the four XIS detectors measure a systematically higher flux with respect to the pn, especially in the soft energy range. However we note that the cross-calibration accuracy between the pn and the XIS is, especially above 4 keV, of the same order of the discrepancies between the different XIS units.\\
\begin{table}
\begin{minipage}[h]{\columnwidth}
\caption{Normalization factors for the \suz/XIS cameras with respect to the \xmm/pn (see Section \ref{xmm} for details).}
\label{crossc}
\centering
\begin{tabular}{ccccc}
\hline\hline
 Detector & \multicolumn{4}{c}{Energy range} \\
& 1.5--12 keV & 1.5--4 keV & 4--8 keV & 8--12 keV\\
\hline
XIS\,0 & $1.11\pm0.02$ & $1.23\pm0.04$ & $1.05\pm0.03$ & $1.1\pm0.1$ \\
XIS\,1 & $1.08\pm0.02$ & $1.19\pm0.04$ & $1.00\pm0.03$ & $1.1\pm0.2$ \\
XIS\,2 & $1.17\pm0.02$ & $1.26\pm0.04$ & $1.13\pm0.03$ & $1.1\pm0.1$ \\
XIS\,3 & $1.19\pm0.02$ & $1.33\pm0.04$ & $1.12\pm0.03$ & $1.1\pm0.1$ \\
\hline
\end{tabular}
\end{minipage}
\end{table}
\indent To derive the period of \src\ we used the pn data. Photon arrival times were converted to the Solar System barycenter using the \texttt{SAS} task \texttt{barycen}. With a standard folding analysis of the light curves, we measured a spin period of \mbox{$7.5891\pm0.0002$ s}. The resulting peak-to-trough pulsed fraction\footnote{Note that here  we use a different definition of the pulsed fraction than in \citet{met07}. Using the old definition based on a sinusoidal fit to the profile, the estimated pulsed fraction in the same energy range is ($8\pm1$)\%.} \mbox{$PF_{\rm{pt}}\equiv(F_{\rm{max}}-F_{\rm{min}})/(F_{\rm{max}}+F_{\rm{min}})$}, where $F_{\rm{max}}$ and $F_{\rm{min}}$ are the observed background-subtracted count rates at the peak and at the minimum, is $(11\pm2)$\% in the 2--10 keV band.
\section{\int\ observations and broad-band spectral analysis}\label{igr-analysis}
\begin{figure}[h!]
\resizebox{\hsize}{!}{\includegraphics[angle=-90]{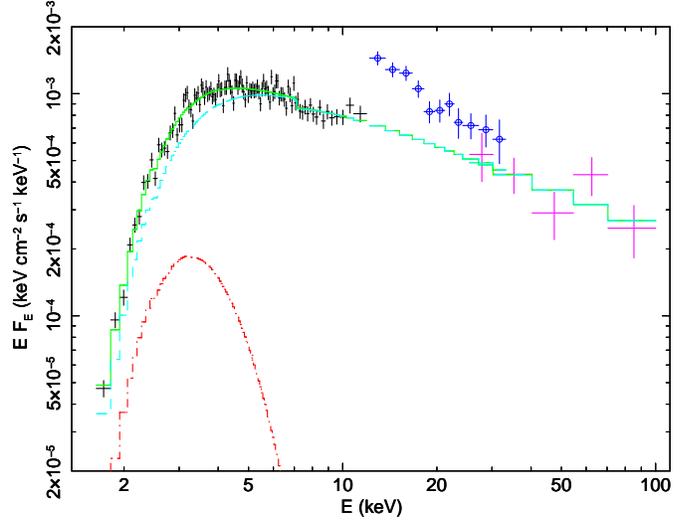}}
\caption{\xmm\ and \int\ broad-band spectrum of \src\ (see the online edition of the article for a color version of this figure). The data from \xmm/pn (black) and \int/IBIS (lilac) are fit with the power-law  (light blu dashed line) plus blackbody (red  dot-dashed line) model. We also plotted \suz/HXD-PIN data using the blu circle marks.}
\label{broad}
\end{figure}

We analyzed the \int's AO-4 Key Project (KP) observation of the Galactic Centre, and report here results obtained with ISGRI, the low energy detector plane \citep[\mbox{15 keV}--\mbox{1 MeV};][]{lebrun03} of \int's imager IBIS\footnote{The direction of \src\ was scarsely covered by \int's X-ray monitor JEM-X \citep{lund03}, due to its smaller (with respect to that of IBIS) field of view (7\degr\ diameter). As a consequence, only about 200 ks of exposure were available, and this was unsufficient for a detection.} \citep{ubertini03}. 
The IBIS/ISGRI data were reduced using the Off-line Scientific Analysis
package (\texttt{OSA} version 6.0). Thanks to IBIS large field of view ($29^{\circ}\times29^{\circ}$), \src, which is located $\sim$10$^{\circ}$ from the Galactic Centre, was almost constantly observed during the KP. This project was divided in two parts: the first one lasted from September 12 to October 5, 2006, yielding an effective exposure time of $\sim$750 ks on the source, while the second one lasted from February 28 to March 25, 2007 for an exposure of $\sim$550 ks.
Our data set consists of about 700 individual pointings, but due to the faintness of the source we could not extract the spectra from the individual pointings. We extracted the images for each pointing in 10 energy bands between 20 and 300 keV and  added all the individual images in order to produce two mosaics (one for each part of the KP observation). We then extracted the fluxes from the mosaics in each energy band in order to derive the source spectrum, and rebinned the ISGRI response matrix to match our energy bands. Since the fluxes were found not to vary within the errors between the two parts of the KP, we added the two spectra in order to increase our statistics.\\
\indent The IBIS/ISGRI spectrum can be well fit ($\chi^2_r=0.85$ for 5 d.o.f.) with a single power-law, with $\Gamma= 1.7\pm0.3$, and a \mbox{20--60 keV} flux of \mbox{$(2.8\pm0.4)\times10^{-11}$ \flux}.
The power-law parameters obtained in the hard X-ray band with IBIS/ISGRI and the \suz/HXD (Section \ref{hxd-analysis}) are consistent within the errors. The joint fit of the IBIS/ISGRI and \suz's XIS and HXD-PIN data to the broken power-law plus blackbody model adopted in Section\,\ref{hxd-analysis} yields virtually identical best-fit parameters, with $\chi^2_r=1.09$ for 358 d.o.f..\\
\indent We also fit the IBIS/ISGRI spectrum simultaneously with the \xmm/pn spectrum described in the previous section, using a blackbody plus power-law model. The resulting best-fit parameters ($\Gamma=1.55\pm0.08$, blackbody temperature \mbox{$k_BT=0.6\pm0.1$ keV} and radius \mbox{$R_{\rm{BB}}=2^{+2}_{-1}$ km} (at 15 kpc), and \mbox{$N_{\rm H}=6.6^{+0.5}_{-0.4}\times10^{22}$ $\rm cm^{-2}$}, with $\chi^2_r=1.27$ for 107 d.o.f.) are consistent with an extrapolation of the low-energy model (Fig.\,\ref{broad}).

\section{Discussion}
The \suz, \xmm, and \int\ observations reported here represent a
complementary data set that allows us to study the spectral
properties of \src\ in the broad \mbox{1--100 keV} energy range.
Although the  \suz/HXD does not have imaging capabilities, we know
thanks to  \int\ that no other bright hard X-ray sources are
present in its field of view.  The uncertainties in the
instrumental background (currently at the $\sim$5\% level) and in
the modelling of the Galactic Ridge emission are a more relevant concern.
Future improvements in the knowledge of these components may
eventually allow us to obtain more robust conclusions. Thanks to
its imaging capabilities, background issues do not affect the
\int\ observations. However, the IBIS/ISGRI data required long
integration times, with discontinuous observations spanning
several months. Thus they provide information on the average
properties only. The possible presence of long term variability
was in fact one of the main motivations to perform the \suz\ and
\xmm\ observations simultaneously. With all these caveats in mind
we proceed
now to discuss the broad band spectrum of \src.\\
\indent The \xmm\ and \int\ spectra are consistent with an extrapolation of the power-law plus blackbody model measured in the 2--10 keV band.
Between 12 and 30 keV the \suz/HXD sensitivity is better than that of \int, allowing to detect \src\ during a single 50 ks long observation. With respect to the \xmm\ and \int\ joint fit, the HXD data show an ``excess'' (see Fig.\,\ref{broad}) that cannot be completely ascribed to calibration uncertainties between the various instruments.\\
\indent Given the lack of a direct measure of the Galactic Ridge emission around \src, we cannot exclude that this excess is due to an underestimation of such contribution to the background. If instead the excess is a real feature of the spectrum of \src, its broadband spectrum could be empirically modeled adopting a power-law with the photon-index changing from $\sim$1 to $\sim$2 at $\sim$16 keV, and a blackbody component with $k_BT\sim0.8$ keV. This would agree with the results reported by \citet{gotz06}, who point out that the hard tails of the SGRs are softer than the power-law components measured below \mbox{10 keV}.\\
\indent The presence of a down-break in the 10--20 keV spectrum of \src\ would have remarkable physical implications. The soft X-ray emission from magnetar candidates (SGRs and AXPs) is usually interpreted within the twisted magnetosphere model as due to resonant cyclotron up-scattering of soft photons from the star surface onto charges flowing into the magnetosphere \citep{tlk02}. Detailed calculations of resonant compton scattering (RCS) spectra have been recently performed \citep{lyutikov06,fernandez07} and successfully applied to fit AXP spectra \citep{rzl07}. Quite interestingly, some of the model spectra presented by \citet{fernandez07} exhibit a downward break in the tens of keV range. Their overall shape is quite reminiscent of the \suz\ XIS/HXD-PIN spectrum of \src, and, as noted by \citet{fernandez07}, they may also play a role in the interpretation of the broadband X-ray spectrum of SGR\,1900+14. In particular, when assuming a (non-thermal) top-hat or a broadband velocity distribution for the magnetospheric charges, multiple peaks can appear in the spectrum (see their Fig.\,6 and Fig.\,11). The downturn possibly present in our data may then be due the presence of a second ``hump'' (in addition to the main thermal one) in the range \mbox{10--20~keV}. Nobili, Turolla, \& Zane (in preparation) assuming a 1-D thermal electron distribution superimposed to a (constant) bulk velocity, found also double humped spectra. In this case the second (and only) hump occurs when resonant scattering is efficient enough to fill the Wien peak at the temperature of the comptonising particles. A spectral break at $\sim$15~keV would translate then in a temperature of $\sim$5 keV for the magnetospheric electrons. If a more refined treatment of background subtraction confirms the spectral break in the X-ray data of \src\, this would provide important diagnostics for the physical parameters of the model.\\
\indent In 2003 we started a long-term monitoring program to study the time evolution of the spectral properties of \src\ using the \xmm\ X-ray satellite. The December 27, 2004 giant flare was a fortunate occurrence that allowed us to observe how the source properties evolved in the two years leading up to the flare and how they changed after this dramatic event \citep[see][for details]{mte05,met07,rea05_atel,tiengo05}. The \xmm\ data showed a doubling of the flux in the September--October 2004 followed by a gradual recovery to the ``historical'' level after the giant flare. A direct comparison of the \xmm/pn count rates measured in the different observations shows that before the giant flare the flux of \src\ in the \mbox{1--10 keV} band was monotonically increasing, while the three observations after the flare, and preceding the one reported here, followed a steady decreasing trend. The September 2006 observation breaks this long term decay, having a count rate higher by 5\% with respect to the last \xmm\ observation performed 5 months before. This slight (but statistically significant) re-brightening might indicate either a temporary oscillation around an equilibrium flux level or the start of a  new monotonic flux increase, similar to the one that preceded the December 2004 giant flare. This phenomenon was interpreted as due to the building up of a magnetospheric twisting, that determined also the hardening of the X-ray spectrum, an increase of the spin-down and a more intense bursting activity \citep{mte05}. The relatively high burst rate observed during the \xmm\ and \suz\ observations of September 2006 (see Section\,\ref{burstingemission}) is therefore another indication of a possible increase of the magnetospheric twisting in \src, but, before a new \xmm\ observation will be performed, only the monitoring of the frequency and intensity of \src\ bursts can tell us if the evolution is erratic or follows a stable trend. The recent report of a bright burst from \src\ \citep{golenetskii07,perotti07} seems actually to favour the second hypothesis.\\
\indent Two of the bursts detected during the \suz\ observation were bright enough to allow spectral analysis. In both cases, the broadband spectrum (\mbox{2--100 keV}) revealed the presence of two components: a soft component which is well reproduced by a blackbody with $k_BT \sim 2$--4~keV, and a harder one whose
spectral shape is not firmly established and can be equally well fit with a power-law, a hot bremsstrahlung or a second blackbody (See Table\,\ref{bfits} and Fig.\,\ref{burstfits}). In absence of robust theoretical predictions, we can not exclude that a two component model simply reflects our ignorance of the correct spectral shape, and has therefore a purely phenomenological significance. However, it is worth noticing that, from their recent analysis of a sample of 50 bursts detected from \src\ with \hete\ from 2001 to 2005, \citet{nakagawa07} have suggested the presence of a time delay between the 30--100~keV and the 2--10~keV emission. Although such a delay can be attributed to an intrinsic, rapid spectral softening,
an alternative, and simpler, interpretation invokes the presence of two separate emitting regions.\\
\indent Let us consider a scenario in which the two
components are physically distinct and let us consider the hard component first.
In the magnetar scenario, short bursts are usually ascribed to either reconnection
phenomena in the external magnetosphere \citep[eventually modulated by a tearing instability, see][]{lyutikov03} or movements
of the footprints of the external magnetic field, produced by crustal deformations
or fractures driven by the stress exerted by the internal field
helicity \citep{thompson95,thompson01,tlk02}.
Both kind of processes lead to the generation and launch of an Alfv\`en wave, which produces and accelerates a particle cascade, and ultimately is detected as a burst.
The emerging spectrum is expected to be synchrotron dominated, unless the Alfv\`en wave is temporarily trapped in a fireball and thermalized. Therefore, both the \mbox{BB + PL} and BB + BB spectral fits are consistent with a scenario in which such an Alfv\`en wave is responsible for the hard component. We notice that, although a fireball formation is not required to explain short bursts (and therefore usually not invoked in such cases), to our knowledge there is no a priori reasons why a small fireball can not be created and evaporated in a sub-second time scale, giving rise to a thermal spectrum. A point in favour of this interpretation is that, in the BB + BB fit, the temperature of the hot blackbody is remarkably close to the minimal temperature above which a fireball thermalizes, $k_BT \approx 11\,(R_{\rm{NS}}/10\,\rm{km})^{-1/5}$~keV, according to \citet{thompson01}, eq. [71] \citep[see also][for similar findings based on longer duration burst]{olive04}. The third model of the hard component which is compatible with our data is a bremsstrahlung emission at $\sim$100~keV. Quite recently, \citet{thompson05} and \citet{beloborodov07} discussed the electrodynamics of the magnetar coronae and the production mechanisms for soft gamma-rays. In particular, their model predicts the existence of a thin transition layer between the corona and the thermal photosphere, where Langmuir turbulence can be excited by a downward beam of current-carrying charges. As a result, the transition layer can be heated up to a typical temperature of $\sim$100~keV, and emit,  approximatively, an optically thin bremsstrahlung at a single temperature. Although \cite{thompson05} model was originally developed in connection with the persistent hard emission of magnetars, the predicted bremsstrahlung temperatures are remarkably close to those we detected during the two bursts, suggesting that a similar mechanism may instead be activated during periods of activity.\\
\indent Our results about the spectral modelling of the soft X-ray component are more robust, inasmuch as all our spectral fits require the presence of a cold blackbody with \mbox{$k_BT \sim 2$--4~keV}. This is in agreement with similar findings by \cite{olive04}, \cite{nakagawa07}, and \citet{feroci04}. This component is usually interpreted as due to emission from a fraction of the star surface (which can be as large as the whole star in the case of our BB + BB fit) heated by returning currents. Alternatively, it has been suggested that the soft component may originate up in the magnetosphere ($\leq$700~km), presumably due to a delayed emission process \citep{nakagawa07}. Here we only notice that, although the spectra of our two events are compatible with emission from the star surface, the radius of the cold blackbody as measured during other short bursts can reach values much higher than 50--100~km \citep[][similar findings have been found in the case of SGR\,1900+14 bursts measured with \emph{Swift}, Israel et al. (in preparation)]{nakagawa07}. One possible explanation is that part of the flare energy is intercepted and reprocessed in a larger region and re-emitted at a lower temperature. In such scenario, the radius of the reprocessing region can then vary depending on the fraction of material that is intercepted, and is not bounded by the value of the star radius. \citet{thompson01} considered the equilibrium state of a pair corona sufficiently extended that the local value of the magnetic field is $B\ll B_{\rm{QED}}$, so that photon splitting can be ignored (if the magnetic field scales as a dipole, this occurs above $\sim$3 star radii, for a polar surface value of $\sim$$10^{15}$~G). They found that a stable balance between heating and diffusive cooling requires a continuous source of ordinary photons that can be provided, for instance, by external illumination. If the corona intercepts part of the flare beam (which in their treatment is assumed to originate in a trapped fireball, although in general this is not necessarily required), equilibrium is possible below a critical luminosity given by
\begin{displaymath}
L<L_{\rm{max}} = 1.5 \times 10^{42}\, \tau_{EO}^{-1} \left ( \frac {k_BT_e} {20\ {\rm
keV} } \right )^4 \left ( \frac {R }{10\ {\rm km}} \right )^2 \, {\rm
erg\ s^{-1}} \, ,
\end{displaymath}
where $T_e$ is the pair temperature in the corona, $\tau_{EO} \sim 1$ is the scattering depth for ordinary to extraordinary mode conversion and $R$ the radius of the emitting part of the corona (see equations [84] and [89] in \cite{thompson01} and note that there is a typo in their equation [89]: it should contain $R^2$ instead of $R^{-2}$). Even for an emission region as small as 5--20~km (as inferred by our best fit of the low temperature blackbody) and a temperature of $k_BT \sim 2$--4~keV, this is well above the luminosity emitted during the two bursts detected by \suz. Therefore, simply on the basis of energetics, relatively large emitting regions for the cold blackbody are compatible with the temporary formation of a pair corona, sustained by a fraction of the flare energy. When the heating rate ceases, the pair atmosphere contracts and quickly evaporates. In order to derive firmer conclusions a more detailed analysis is needed, mainly in assessing the possibility that the intercepted beam is thermalized and re-emitted as a
blackbody. This study is beyond the purpose of this paper, and will be presented elsewhere (Israel et al. 2007, in preparation).

\begin{acknowledgements}
This work is based on data from observations with Suzaku, INTEGRAL, and XMM-Newton.
Suzaku is a Japan's mission developed at the Institute of Space and Astronautical Science of Japan Aerospace Exploration Agency in collaboration with U.S. (NASA/GSFC, MIT) and Japanese institutions. We would like to thank the Suzaku operations team for their support.
XMM-Newton is an ESA science mission with instruments and contributions directly funded by ESA Member States and NASA.
INTEGRAL is an ESA project with instruments and science data centre funded by ESA member states (especially the PI countries: Denmark, France, Germany, Italy, Switzerland, Spain), Czech Republic and Poland, and with the participation of Russia and the USA.
This research has made use of HEASARC online services, supported by NASA/GSFC, and of the SIMBAD database, operated at CDS, Strasbourg, France.
The Italian authors acknowledge the support of the Italian Space Agency (contracts I/023/05/0 and I/008/07/0) and the Italian Ministry for University and Research (grant PRIN 2005 02 5417).
SZ acknowledges STFC for support through an Advanced Fellowship.
DG acknowledges the French Space Agency (CNES) for financial support.
NR is supported by an NWO Post-doctoral Fellowship. 
Finally, we wish to thank the anonymous referee for providing useful comments on the manuscript.
\end{acknowledgements}
\bibliographystyle{aa}
\bibliography{biblio}
\end{document}